# Range of Motion Sensors for Monitoring Recovery of Total Knee Arthroplasty

Minh Cao[1], Brett Bailey[1], Wenhao Zhang[2], Solana Fernandez[1], Aaron Han[1], Smiti Narayanan[1], Shrineel Patel[1], Steven Saletta[1], Alexandra Stavrakis[3], Stephen Speicher[1], Stephanie Seidlits[1], Arash Naeim[2], Ramin Ramezani[2]

1. Bioengineering Dept, University of California, Los Angeles, Los Angeles CA 90095, USA
2. Center for Smart Health, University of California, Los Angeles, Los Angeles CA 90095, USA
3. Orthopaedic Surgery, University of California, Los Angeles, Los Angeles CA 90095, USA

{mcao21, brettrbailey, wenhaoz, sbfernandez, seidlits, raminr}@ucla.edu, {AStavrakis, anaeim}@mednet.ucla.edu, {caaron115, smitin9, shrineelpate01, smsdodgers, speicher.stephen}@gmail.com

*Abstract*— A low-cost, accurate device to measure and record knee range of motion (ROM) is of the essential need to improve confidence in at-home rehabilitation. It is to reduce hospital stay duration and overall medical cost after Total Knee Arthroplasty (TKA) procedures. The shift in Medicare funding from pay-as-you-go to the Bundled Payments for Care Improvement (BPCI) has created a push towards at-home care over extended hospital stays. It has heavily affected TKA patients, who typically undergo physical therapy at the clinic after the procedure to ensure full recovery of ROM. In this paper, we use accelerometers to create a ROM sensor that can be integrated into the post-operative surgical dressing, so that the cost of the sensors can be included in the bundled payments. In this paper, we demonstrate the efficacy of our method in comparison to the baseline computer vision method. Our results suggest that calculating angular displacement from accelerometer sensors demonstrates accurate ROM recordings under both stationary and walking conditions. The device would keep track of angle measurements and alert the patient when certain angle thresholds have been crossed, allowing patients to recover safely at home instead of going to multiple physical therapy sessions. The affordability of our sensor makes it more accessible to patients in need.

*Keywords— arthroplasty, accelerometer, range of motion*

## I. INTRODUCTION

After total knee arthroplasty (TKA), it is crucial for patients to undergo physical therapy and bending of the knee to allow for proper healing and to prevent arthrofibrosis – defined as excess scar tissue formation – which reduces range of motion (ROM) at the knee [1]. During the recovery process, a patient may be prescribed to extend their knee to a certain angle during physical therapy exercises and daily use. Movement of the knee in the 1-2 weeks after TKA is essential in maximizing ROM retention and it is reported that moderate to severe stiffness after surgery – less than 90° flexion – occurs in 3.7% of patients undergoing post-operative recovery of TKA [2]. While ROM is the most descriptive variable in final flexion after TKA, existing knee angle calculation methods are inaccurate and expensive [3] [4].

Traditionally, patients would undergo post-operative recovery and physical therapy at a clinic and were billed following a pay-for-service model. However, the new Bundled Payments for Care Improvement (BPCI) initiative in the United States has changed Medicare funding to a bundled payment model; instead of being billed for individual services, patients are charged a lump sum up-front for all costs [5]. Integrating a ROM sensor system into the dressing will allow surgeons to charge Medicare for the cost of the sensor in the bundled payment [6]. Therefore, patients will not have to pay out of pocket for the cost of the post-operative recovery sensors like in the current system. This new initiative has resulted in sending patients out of the hospital and back home for recovery [7]. Patient compliance with recovery exercises at home is difficult to ensure. They may not perform exercises correctly: they may overextend or not flex their knee as recommended by physicians. Therefore, there is a need for a device that can monitor patients' at-home compliance with recovery activities, while providing them with qualitative and quantitative feedback.

The advent of Internet of Things and cloud computing have greatly empowered the sensor-based data collection and data transfer in an affordable and efficient manner [16]. At-home remote healthcare monitoring systems, equipped with low-cost wearable sensors, have been proposed to track the individual movements, aiming to improve the quality of at-home care [17] [18]. There exist remote patient monitoring systems that infer classification of human movements using a 3-axial accelerometer. Such systems have demonstrated the capability of providing valuable insights to the physicians about the patient's recovery process by analyzing the changes in their physical activities [19][20].

The current methods to track knee ROM are goniometers and visual tracking systems [1]. Goniometers - while cheap and convenient - require training and visual estimation of bending angle, often leading to inaccurate measurements [8]. Additionally, they cannot be used by patients to monitor ROM during their daily activities. Visual tracking system on the other hand is much more accurate, however, it requires the patient to set up camera systems that are expensive and immobile [9][10]. Other devices on the market, currently, provide a host of capabilities but are expensive and significantly lack the ability to function as ROM sensors. The Smart Knee allows for accurate recording of the flexion of the knee [11]. However, it is expensive and relies on attachment to the lateral side of the leg, preventing integration into the post-operative surgical dressing. The APDM Opal is capable of extensive motion analysis, but its cost, size, and inability of knee ROM sensing make it difficult to incorporate into the recovery stage of TKA [12].

To overcome these challenges, we developed a ROM sensor that uses two accelerometers to calculate the angle of knee bending. The accelerometers measure the acceleration at the knee joint which is used to calculate the tilt angles in the x, y, and z planes. It then calculates knee angle during stationary and ambulatory positions. Our proposed system can be integrated into the post-operative dressing where to the knowledge of the authors no other sensor available for commercial purchase is currently capable of. The sensors would be attached to the proximal and distal ends of the dressing.

## II. Materials and Methods

Absolute knee angle of bending is calculated using two commercially available Sony Smartwatch 3 with built-in EM7180 $\pm$ 2 g triaxial accelerometer, placed on the anterior thigh and shank, respectively. The Sony Watch 3 acts as our prototype on body accelerometer sensor at 250 Hz sampling rate, but any smartwatch with accelerometer can be used in similar manner for prototyping. The watches are positioned in a way that the watch faces are flush to the thigh and shank (Figure 1). All the methods presented in this paper will be done according to this orientation.

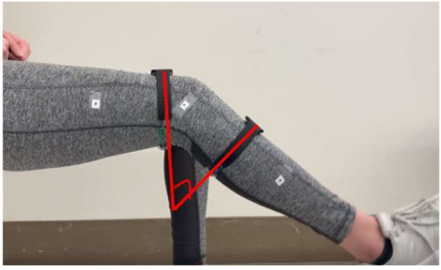

Fig. 1. Starting position of sensors and the desired angle calculated.

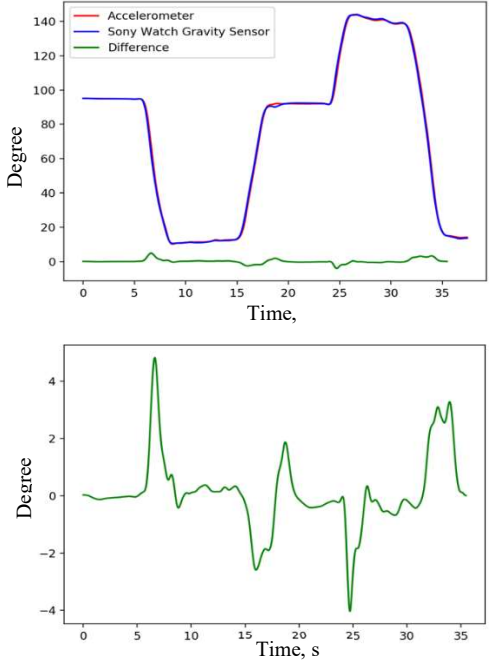

Fig. 2. (top) the angles calculated by the different sensors (bot) difference in angle calculation measured by the gravity sensor and accelerometer on the SmartWatch.

The accelerometer data collected from the watches at 250 Hz will be comprised of two main components: 1) the static gravity acceleration downward from the world Z axis, and 2) the linear acceleration due to movement of the watches. For angle calculation purpose, we are only interested in the static gravity component. To isolate this component, we pass the accelerometer data through a fourth-order Butterworth low-pass filter with a 1 Hz cutoff. This method can be used to emulate a gravity sensor, which measures the contribution of Earth's gravitational acceleration on the device (Figure 2).

The accelerometer sampling frequency was set at 250 Hz. Acceleration due to gravity calculated measurements on the Sony SmartWatch 3 were used as a baseline to test the accuracy of the low-passed accelerometer data. The motivation for using the accelerometers instead of relying on the Sony Watch internal sensor calculations was that tri-axis accelerometers are cheap and widely available, while Sony Watches are bulky and expensive. Smartwatches are used in this experiment due to the convenient data collection and storage.

### A. Method of Angle Calculation: Difference from World Z using Inverse Cosine

In this method, the angle of bending of the knee is derived from the difference of the absolute angle of each of the watches that are calculated independently. The absolute angle of the watches with regarding to world Z axis is determine by the proportion of the gravity component magnitude on the Z axis compared to the total gravity component in all axes. This is because although the total gravity component for all axis remain constant at all time, the gravity component on Z axis shift toward the X and Y axis as the knee bends upward and vice versa. The cosine of this proportion will give the absolute angel of the watch.

In these equations, $Z_{device}$ is the acceleration due to gravity felt in the Z direction of the sensor. and $SM_{X,Y,Z}$ is the signal magnitude of the accelerations due to gravity in the X, Y, and Z axes of the sensor. This $SM_{X,Y,Z}$ should be approximately equal to the positive value of acceleration due to gravity, 9.81 m/s$^2$, after filtering.

$$\cos(\theta_z) = \frac{Z_{device}}{SM_{X,Y,Z}} \quad (1)$$

$$\theta_z = \cos^{-1}\left(\frac{Z_{device}}{SM_{X,Y,Z}}\right) \quad (2)$$

The angle of bending ($\theta_d$) is calculated using Eq. 3 (Figure 3).

$$\theta_d = \theta_{bottom} - \theta_{top} \quad (3)$$

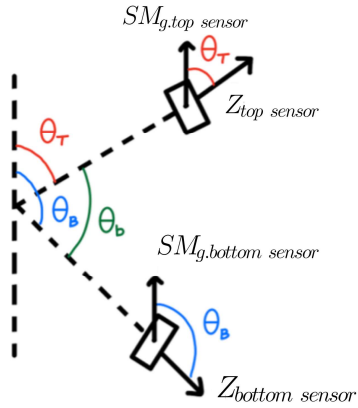

Fig. 3. Angle between the sensors ($\theta_b$) calculated as $\theta_B - \theta_T$

The knee bending exercise can be modeled as an uni-axial joint. In an ideal scenario where both the X and Y planes of the two mounted accelerometer are aligned, as the knee bends, only the displacement in the Z angle contributes to the bending motion. However, in real world, the X and Y axis of the two accelerometers will not align due to natural movement, slipping off the mounted sensors as well as twitching motion during bending exercise. Therefore, error correction for misalignment in the X and Y axis is crucial.

For X axis misalignment correction, the default X angle of each device will be $\theta_z+90°$. Calculated angle values will always be positive for all axes, since the range of $\cos^{-1}$ is 0° to 180° (Figure 4). When the X angle is less than 90°, it correlates to the watch sensor tilted to the opposite side with respect to the vertical direction. In this case, the measured Z angle is multiplied by -1 (Figure 5).

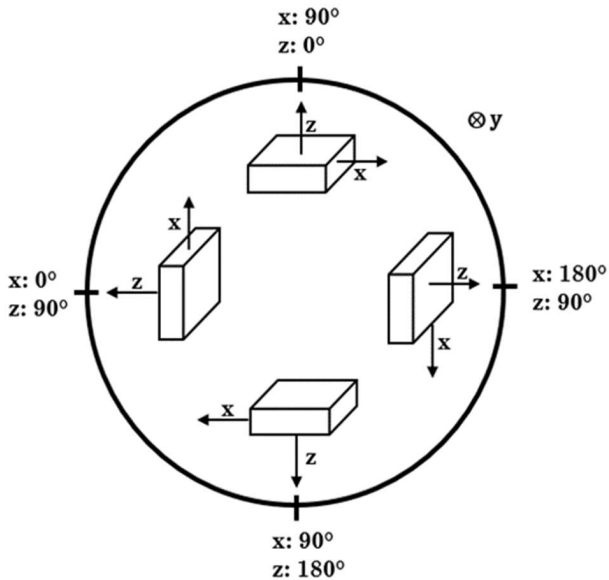

Fig. 4. Rotation of SmartWatch and subsequent codependent X and Z angle recordings in the X-Z plane

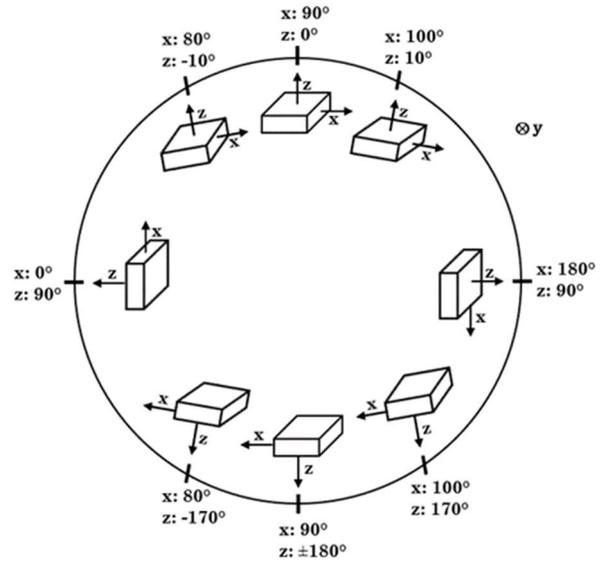

Fig. 5. Rotation of device and subsequent X and Z angle recordings with applied inversion.

For Y axis misalignment correction, the contribution of the Y angular offset component of gravity is scaled to zero as the Z angle approaches 90° (acceleration due to gravity in the Z axis approaches zero) and can be expressed as:

$$\theta_{Z,f} = \theta_{Z,i} - \left( |90 - |\theta_Y|| \times \frac{90 - |90 - \theta_x|}{90} \right) \quad (4)$$

In Eq. 4, the Z angle is multiplied by a scaling factor which is equal to 1 at Z = 0° (vertical) and 0 at Z = 90° (perpendicular to acceleration due to gravity). The component of $\theta_Y$ that should be added to $\theta_Z$ is $|90 - |\theta_Y||$ since $\theta_Y$ ranges from 0° to 180° and is 90° when there is no Y-offset. This component is scaled, since the contribution of Y in Z angle decreases as Z rotates toward the X-Y plane. When the Z angle is aligned in the X-Y plane (perpendicular to acceleration due to gravity), it is unaffected by Y-offsets. This is why the scaling factor must be 0 when $\theta_Z = 90°$.

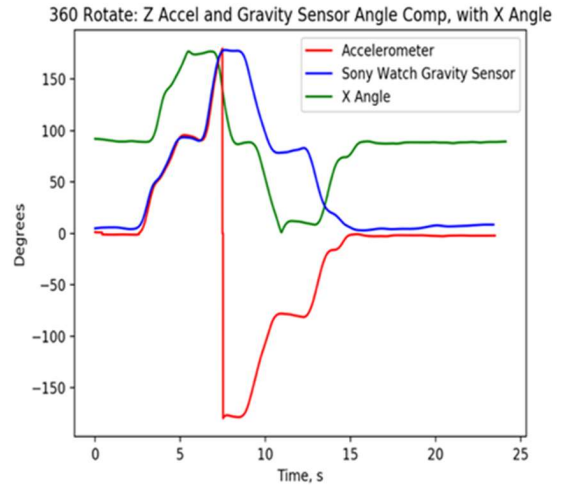

For example, if we take the angle measurements from Figure 7 (x, y, z) = (90, 70, 20) and according to Eq. 4, we get 0 degrees difference. This is the desired result for the Z angle during this phase of the Y-offset test. To demonstrate the correction factor at ~45 degrees, we will consider another point in time during the test. At t = 35 seconds, the knee was bent at ~55 degrees and the recorded angles were (150, 125, 70). Applying the correction factor the system arrives at the degree difference of 58.

$$\theta_{Z,f} = 70 - \left(|90 - |125|| \times \frac{90 - |90 - 150|}{90}\right) = 58$$

This is the within 3 degrees of the expected result for the Z angle during this phase of the Y-offset test. This is due to the slight offset in Y of the X angle as it rotates out of the X-Y plane. The result of applying this correction factor can be seen in Figure 12.

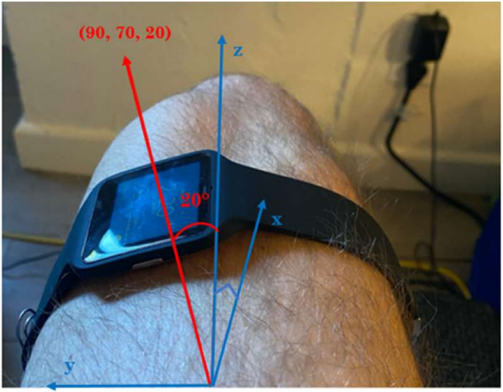

Fig. 7. Y angle offset causing an error in Z angle calculation when the Z angle is at 90°. The recorded angle of the device (x,y,z) = (90, 70, 20).

To test the effectiveness of this scaling factor, the following test was performed. At approximately 0, 55, and 90°, the sensor was offset in the negative Y direction (Figure 8b), then in the positive Y direction (Figure 8c). When the Z angle is 90°, the Y acceleration due to gravity does not change when the devices is rotated about the Y axis. In the Z angle calculation, |90 - θ_X| is used instead of θ_Z since θ_X is less affected by Y fluctuations when Z is close to 0° and gives a more consistent scaling factor.

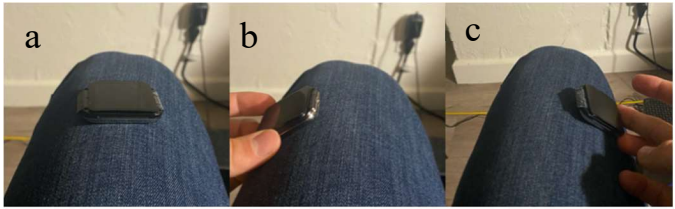

Fig. 8. Positioning of sensor for Y-offset test. (a) no offset (b) offset in the negative Y direction (c) offset in the positive Y direction.

## B. Method of Angle Calculation: Difference from World X-Y Plane using Inverse Tangen

In [15], the knee bending angle is calculated as the angle between the world X-Y plane and the Z direction of the sensor derived from the acceleration due to gravity in the X, Y, and Z directions of the sensor.

$$\theta_z = \tan^{-1}\left(\frac{Z_{device}}{\sqrt{X_{device}^2 + Y_{device}^2}}\right) \quad (5)$$

To calculate difference:

$$\theta_d = \theta_{top} - \theta_{bottom} \quad (6)$$

where $\theta_d$ is the difference in the angle between the thigh sensor and the shank sensor. This method is accurate only when both sensors are on the same side of the Z-axis, which was the case in their Quadriceps Strengthening Mini-Squats (QSM). This way we can only measure an uni-planar angle up to 45 degrees. In real world scenario, this method is not efficient due to the wider range of motion between -5 and 135 degrees.

Our solution is to divide the X-Z plane into four quadrants and assign unique calculations according to the position of the sensor in each quadrant (Figure 8).

TABLE I. CLASSIFICATION OF QUADRANT FOR Z ANGLE IN DECISION ALGORITHM

| Quadrant | X | Z |
|---|---|---|
| I | - | + |
| II | + | + |
| III | + | - |
| IV | - | - |

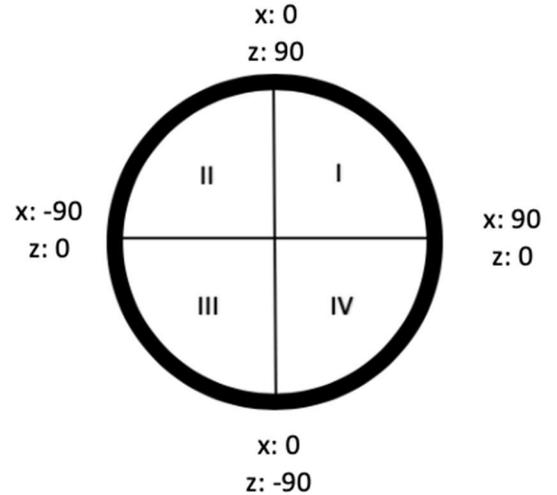

Fig. 9. Angular Determination of Coordinate Grid

If the sensors are on the same side of the x axis (ex. Top: QI, Bottom: QIV), the following calculation is performed:

$$\theta_d = \theta_{top} - \theta_{bottom} \quad (6)$$

If the sensors are on opposite sides of the x axis (ex. Top: QI, Bottom: QIII), the following calculation is performed:

$$\theta_d = 180 - (\theta_{top} + \theta_{bottom}) \quad (7)$$

Another way of visualizing this angle:

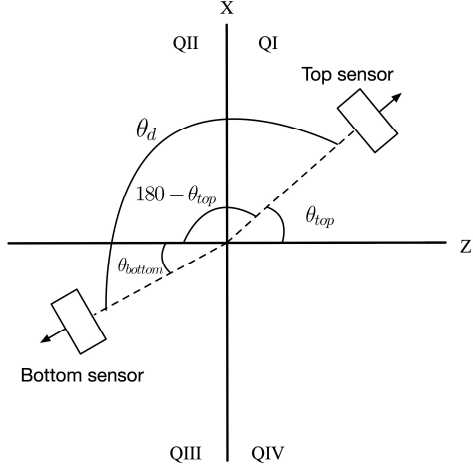

Fig. 10. Angular Determination between sensors using coordinate system.

The Z angle calculations are also needed correction from the Y-axis offset similar in the Method 1. To take into account of the different determination of angle measurement, the following calculation is performed on the top and bottom sensor Z angles to scale by the component of Y in the Z plane:

$$\theta_{Z,f} = \theta_{Z,i} + \left(|\theta_Y| \times \frac{90 - |\theta_X|}{90}\right) \quad (8)$$

In Eq. 8, the Z angle is multiplied by a linear scaling factor which is equal to 1 at Z = 90° (vertical) and 0 at Z = 0° (parallel to X-Y Plane). (90 - $|\theta_X|$) is in the numerator since the y-offset has larger effect when $|\theta_Z|$ is vertical (close to 90 degrees). The component of $\theta_Y$ that should be added to $\theta_Z$ is $|\theta_Y|$ since $\theta_Y$ ranges from -90° to 90° and is 0° when there is no Y-offset. This component is scaled, since the contribution of Y in Z angle increases as Z rotates out of the X-Y plane. When the Z angle is aligned in the X-Y plane, it is unaffected by Y-offsets. This is why the scaling factor must be 0 when $\theta_Z = 0°$.

We demonstrate the effect of this error correction in figure 12. At t = 35 seconds, the recorded angle measurements (x, y, z) are (-52, -35, 20), with the expected angle ~35 degrees. Applying the correction factor (Eq. 8), we get:

$$\theta_{Z,f} = 20 + \left(|-35| * \frac{90 - |-52|}{90}\right)$$

$$\theta_{Z,f} = 20 + \left(35 \times \frac{38}{90}\right)$$

$$\theta_{Z,f} = 20 + 15 = 35 \; degrees$$

We perform the same test as in the Method 1(Figure 8), to test the effectiveness of the Y-axis offset scaling factor. The sensor has zero offset in Figure 8a, offset in the negative Y direction in Figure 8b and offset in the positive Y direction in Figure 8c. In the Z angle calculation, the |90 - $\theta_X$| scaling factor gives a more consistent bending angle since $\theta_X$ is less affected by Y fluctuations when Z is close to 90°.

*C. Discretization*

In our use case, physicians want to know how many times the patient bends their knee in specific range such as 30°, 60°, and 90° during a physical therapy exercise. Therefore, discretization of the data is an important feature so that patients and physicians can easily analyze the data. We created a function that takes the angle difference vector, the correlated time vector, and the desired window size as inputs. The window size is the number of samples taken in a desired window time. The function looks at a sliding window of window size and calculates the standard deviation of the angles. If the standard deviation of the window is less than a set threshold (ex. 1°), then this signifies that the patient has not changed the angle of the knee for a certain period of time. The average of the angle and time values in the window time are recorded. The function then waits for the standard deviation of the window to exceed a certain threshold (ex. 4°), which signifies that the knee angle has been changed significantly. Once the threshold has been exceeded, the sliding window will once again detect a standard deviation of less than 1 degree. The process repeats until the entire angle function has been passed.

*D. Computer Vision*

To establish an accurate ground truth for our experiment, we use the computer vision (CV) utilizing OpenCV and ImageJ libraries to to calculate the angle of the knee from a video. We placed red, green, and blue dots on the knee and recorded the bending while the sensors were gathering measurements. The CV program detects the dots and uses them as anchor for angle measurement. In each trial, the red dot was placed in the middle, closest to the knee, the green dot was placed on the thigh, and the blue dot was placed on the calf. Each color was assigned a minimum and maximum color array $[R_{min}, G_{min}, B_{min}]$ and $[R_{max}, G_{max}, B_{max}]$ which determined which range of RGB values would be classified as red, green, and blue. For every color detection that fell within the predefined range, the program located the midpoint of the detection area. After all the three red, green, blue midpoints are identify, The program calculated the angle between red-to-green line and red-to-blue line as shown in Figure 11.

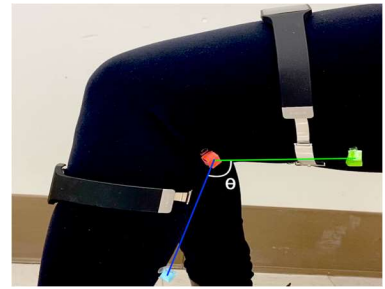

Fig. 11. Lines drawn between red, green, and blue colored dots by the computer vision program. The angle θ between them was calculated using the length of the lines.

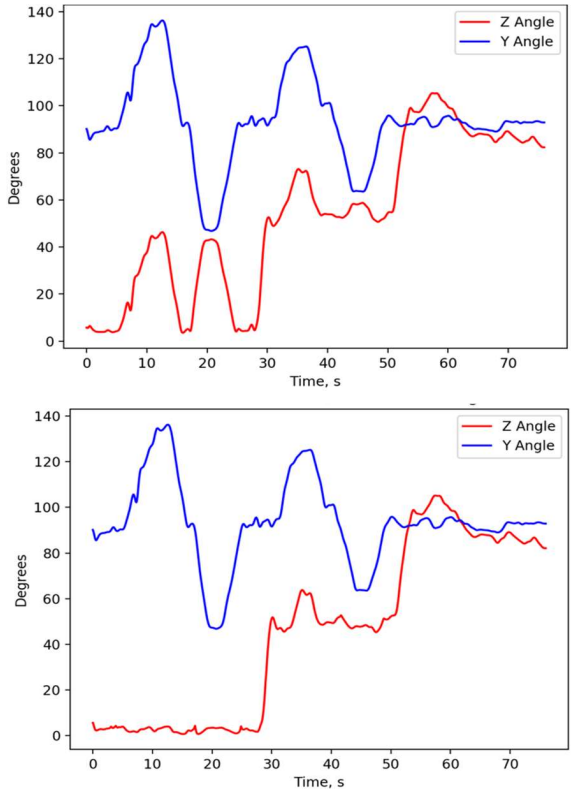

Fig. 12. Method 1 Y-offset test for Z and Y angle. Raw Z angle calculation (top) vs corrected Z angle calculation (bottom) for Method 1. Notice the flattened curve on the Z angle in the bottom figure. This is a result of the correction factor

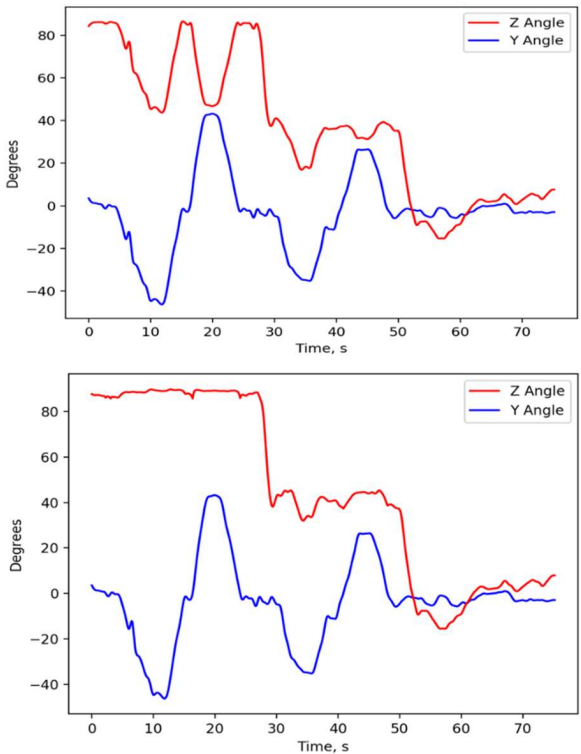

Fig. 13. Raw Z angle calculation (top) vs corrected Z angle calculation (bottom) for Method 2. Notice the flattened curve on the Z angle in the bottom figure. This is a result of the correction factor.

## III. RESULTS AND DISCUSSIONS

The results of the Y-angle test indicate the effectiveness of the scaling factor. Effectiveness is demonstrated by the removal of Z-angle fluctuations induced by Y angular displacement around 0°, 45°, and 90°. Note that there is still slight fluctuation in the corrected graph around 45°, which indicates further room for adjustment of the Z angle correction (Figure 12).

From the Method 2 Y-Angle correction (Figure 13), it is clear that the Z angle is being affected by an offset in the Y direction that must be corrected. By applying Eq 8, the bottom graph is produced. The Y-angle offset is almost entirely eliminated. It is subject to the same fluctuations as the Method 1 correction factor.

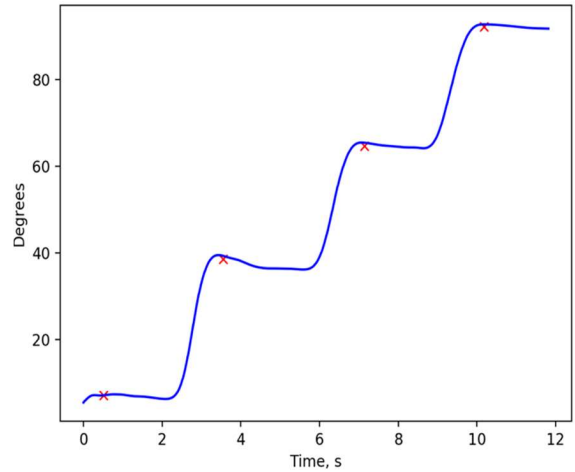

Fig. 14. Discretization with a window size of 1 second applied to a bending test. The red X's indicate the discrete values that the function calculated.

Figure 14 shows this discretization function applied to a bend and hold test where the subject bent their leg to 30°, 60°, and 90° and held each position for about three seconds. During a test, the subject bent their knee from roughly 90° to 0° five times. This test has a maximum error of 20° between the calculated angle using accelerometers and computer vision (Figure 16). For six of the nine peaks, the error is within 5°. The discrepancy between the minimum and maximum error is likely due to inaccuracy in the computer vision program. The computer vison is sensitive to changes in lighting and shadows, which change during knee motion. In the future, we would use a visual tracking system to calculate our baseline, as that would be more accurate. During a walking test, the subject took three steps forward. This walking test shows that our angle calculations can be performed while the knee is moving in multiple axes.

A comparison of *Method A* and *Method B* for our walking test shows the two Methods are within 5° of each other. *Method B* demonstrates closer adherence to the computer baseline curves at the maximum. Further testing with more advanced angle baseline is required before further distinctions between the angles can be concluded. The computer vision angle calculation was performed by on every tenth frame of the walking video. The maximum error between the accelerometer angle calculation for both methods and the computer vision angle calculation is 10° (Figure 17).

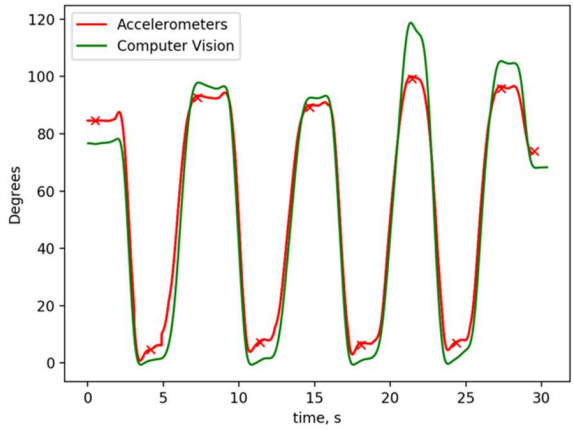

Fig. 15. Bending angle from accelerometers vs computer vision baseline

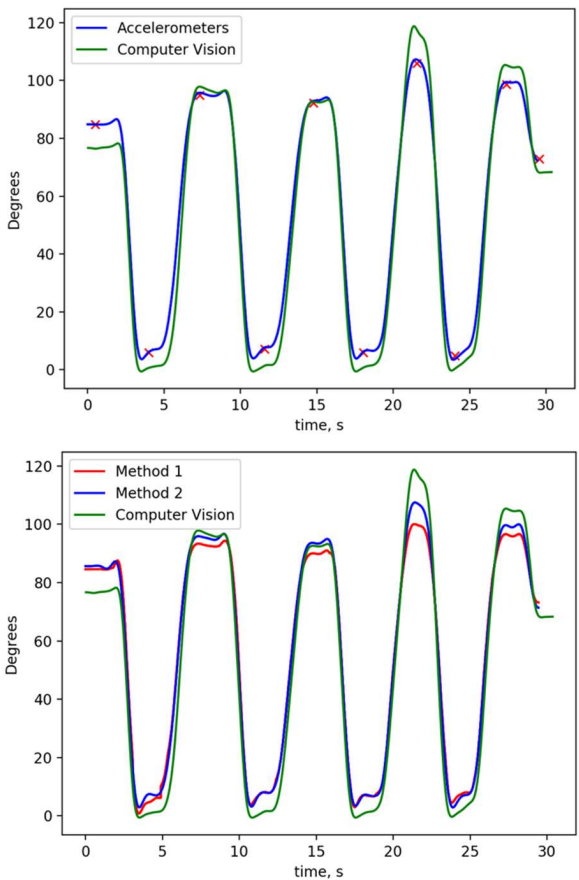

Fig. 16. Angle calculated via Method 1 (top) and Method 2 (Middle) with discretization and a comparison of the two methods during stationary test using accelerometer data against a computer vision baseline.

The computer vision method is limited by the video quality -- blurry frames will result in inconsistent and erroneous angle calculations. Furthermore, if the camera angle is not perpendicular to the sagittal plane of the knee, the warped 2D perspective will yield an incorrect angle. The discrepancy at 3 seconds is likely because there was no image sample taken at that time point when the knee was at its maximum flexion during that step.

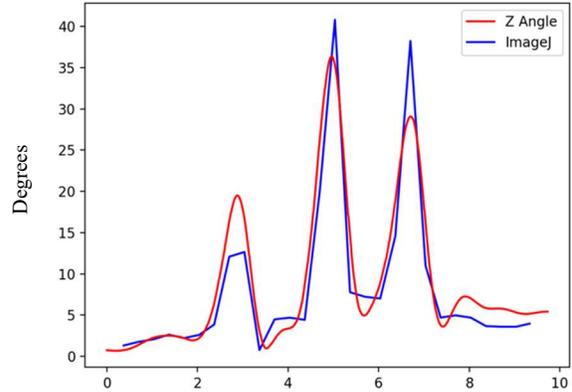

Fig. 17. Comparison of angle calculated via Method 1 during walking test using accelerometer data against the computer vision method baseline.

## IV. CONCLUSIONS AND FUTURE WORKS

In our initial experimentation, the accelerometer data was passed through a low-pass Butterworth filter to isolate acceleration due to gravity and emulate a gravity sensor. Using the accelerometers as gravity sensors, we were able to achieve knee angle measurements in both stationary bending and walking tests. However, establishing a more accurate and precise baseline would be ideal, as computer vision and ImageJ analyses are subject to error due to non-optimal video quality. More complex visual tracking systems that utilize multiple cameras and reflective markers to create a 3D model of the knee are superior in providing reliable angle data. They would provide a more accurate three-dimensional depiction of joints' motions and will therefore better examine the dependability of the knee range of motion device. We would have to ensure that the battery of any deployed device would last 1-2 weeks, which is the lifetime of the post-operative dressing. Additionally, development of an ad-hoc software platform would be beneficial to provide feedback to the patients and physicians.